\documentclass[
    ,final            
  ]
  {aipproc}

\layoutstyle{8x11double}

\begin{document}

\title{Quark-hadron duality in lepton scattering off nucleons}


\author{Krzysztof M. Graczyk}{
  address={Institute of Theoretical Physics, Wroc\l aw University,
 pl. M. Borna 9, 50-204, Wroc\l aw, Poland}
}



\begin{abstract}
Quark-hadron (QH) duality in lepton scattering off nucleons is studied with the resonance quark model. It is  shown that in the case of neutrino scattering off an isoscalar target  the duality  is simultaneously observed for charged and neutral currents $xF_1^{\nu N}$, $F_2^{\nu N}$, and $xF_3^{\nu N}$ weak structure functions.

We demonstrate that the QH duality can be useful property for modeling structure functions in the so-called resonance region. As an example it is shown that combining relativistic quark model predictions with duality arguments allows a construction of the inclusive resonance $F_2^{ep}$ structure function.
\end{abstract}

\maketitle


\section{Introduction}

Historically \textit{quark-hadron duality} was discovered as a remarkable relationship between hadronic scattering amplitudes in $s$ (low energy scale interaction) and $t$ (high energy scale interaction) channels.

The realization of the QH duality in inelastic electron-proton scattering was discovered by Bloom and Gilman \cite{Bloom_and_Gilman} (for review see \cite{Melnitchouk:2005zr}).
It was observed that the $ep$ inelastic  $\nu W_2$ resonance data scales as the deep inelastic
scattering (DIS) one. In practise, it was shown that the resonance $\nu W_2^{ep}  = F_2^{ep}$ data, considered
as a function of $\omega' = 1 + W^2/Q^2$ scaling variable, is averaged by the DIS scaling curve.

The \textit{Bloom-Gilman (BG) duality} constitutes a kind of equivalence between
the resonance and DIS structure functions. As was obtained by De Rujula \textit{et al.} \cite{De Rujula:1976tz} the duality phenomenon can be explained by doing the twist expansion of the $F_2^{ep}$ structure function.
In low-$Q^2$, the leading twist of the $F_2^{ep}$ dominates,
while in high-$Q^2$ higher twist correction become more relevant. Thus the appearance of the BG
duality means the suppression of the contribution from higher twists. De Rujula \textit{et al.} showed also that the Nachtman variable
$\xi(x,Q^2) = 2x/\left(1 + \sqrt{1 + 4 x^2 M^2/ Q^2}\right)
$
is a correct scaling variable to discuss QCD scaling violations, and consequently the proper scaling variable to study the BG duality.

From the practical point of view the duality means equality between two integrals:
\begin{equation}
\label{FESR} \int_{\xi_{min}}^{\xi_{max}} d \xi F^{res}(\xi ,
Q^2_{res}) \approx \int_{\xi_{min}}^{\xi_{max}} d \xi F^{DIS}(\xi,Q^2_{DIS}),
\end{equation}
where  $F^{res}$ and $F^{DIS}$ denote the structure functions at resonance and DIS regions respectively.
The $\xi_{min}$, $\xi_{max}$ are defined by the limits of resonance region, namely:
\begin{equation}
\label{limits}
\xi_{min/max} = \xi(Q_{res}^2, W_{max/min}).
\end{equation}
For the lower bound of the resonance region we take $W_{min}= M+m_\pi$, while $W_{max}$ will vary from $1.6$ to $2.2$~GeV. The typical four-momentum transfer in the resonance region for  $ep$ scattering is $0 < Q^2_{res} < 5$~GeV$^2$. For inclusive $\nu N$ scattering we consider $0 < Q^2_{res} 3$~GeV$^2$. The characteristic four-momentum transfer for DIS is  $Q^2_{DIS}\sim 10$~GeV$^2$.  With such kinematical settings $F^{DIS}$  is integrated over a different kinematical domain, in invariant, mass than $F^{res}$.
\begin{figure}
  \includegraphics[width=1.\textwidth, height=6.5cm]{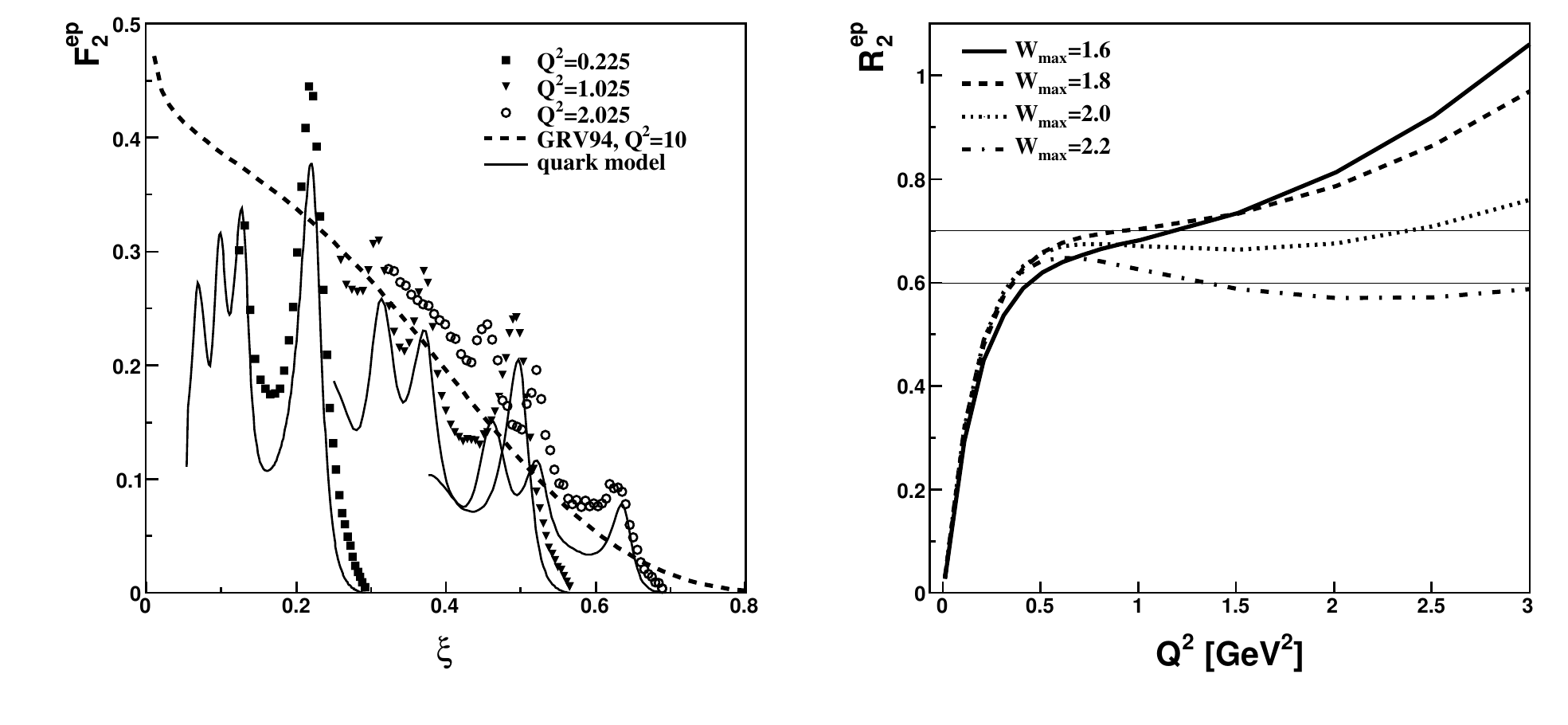}
  \caption{In the left panel, plots of $F_2^{ep}$ structure functions are presented. The
  quark model predictions (solid line)  are compared with the inclusive CLAS data \cite{Osipenko:2003jb} (full squares and triangles and open circles) and DIS
  scaling curve (dashed line).  In the right panel the plots  of the ratios (\ref{ratio_2}), computed for $W_{max}=$1.6, 1.8, 2.0 and 2.0~GeV,  are shown. \label{fig_ep1}}
\end{figure}

Recently the BG duality has been carefully studied by Niculescu {\it et al.} \cite{Niculescu:2000tk}.  The new precision resonance JLab data has been discussed together with the old measurements. The BG duality was quantitatively studied in a wide kinematical region: $0.3< Q^2_{res} < 5$~GeV$^2$, $1.1< W^2 < 4$~GeV$^{2}$. The results of this analysis confirmed the appearance of the BG duality but with accuracy depending on the prominent resonance regions.

The duality is also observed in electron-deuteron scattering \cite{Niculescu:2000tj}, which is also evidence for the duality between electromagnetic neutron structure functions.

It seems that analogously as for the $eN$ scattering, the duality should be observed also for weak nucleon structure functions, which are used to describe the inclusive $\nu N$ interaction. In contrast to the $eN$ scattering the experimental evidence for the duality in $\nu N$ scattering does not exist. The neutrino scattering data, collected in the resonance region, is still too imprecise to order to extract directly the resonance structure functions. On the other hand the knowledge of the DIS structure functions for inclusive $\nu N$ scattering is also limited (see \cite{Morfin_nufact09}).

Given the lack of appropriate neutrino scattering data, the duality in $\nu N$ scattering has been studied only with the phenomenological descriptions \cite{Matsui:2005ns,Graczyk:2005uv,Lalakulich:2006yn}, which have been fine tuned to the old neutrino-deuteron bubble chamber scattering data.

Assuming that the duality is a property of the electromagnetic and weak nucleon structure functions can lead to the additional constraint on: the parameterization of the nucleon and resonance form factors\footnote{In Ref. \cite{De Rujula:1976tz} local duality in elastic $ep$ scattering is discussed. The applications can be found in Ref. \cite{Bodek:2007ym}.} as well as the nucleon inclusive structure functions.

This talk presents an extension of the analysis published in Ref. \cite{Graczyk:2005uv}. The theoretical framework is formulated in Section 2.  In Section 3 we demonstrate that combining quark model predictions and duality arguments leads to the reasonable parametrization of the inclusive $F^{ep}_2$ structure function in the resonance region. The second aim of the talk
is to investigate the BG duality in neutrino scattering off nucleons. In the last section we shortly discuss the duality in lepton scattering off nuclei.
\begin{figure}
  \includegraphics[width=1.\textwidth, height=6.5cm]{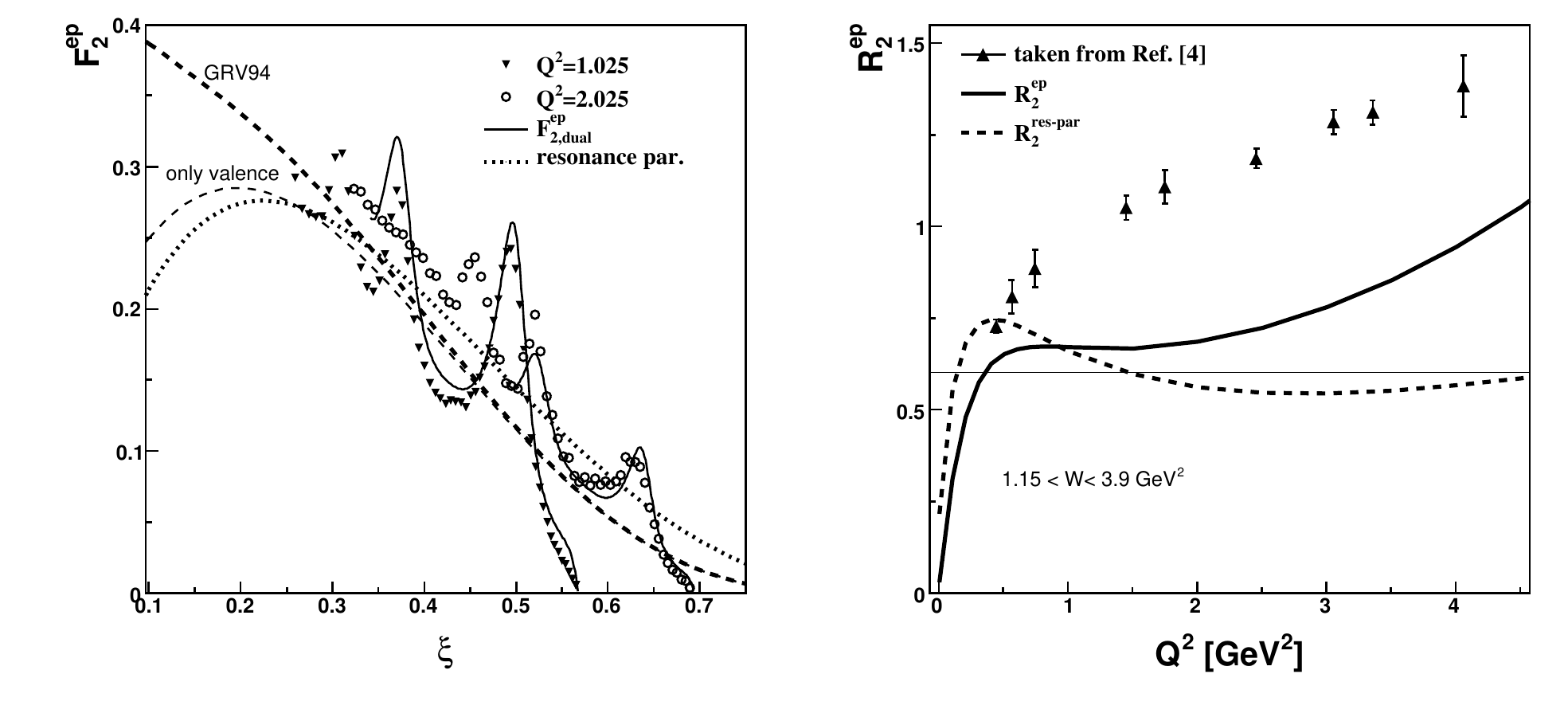}
  \caption{ In the left panel the predictions of the resonance structure function computed with Eq.~\ref{new_F2} (thin solid line) are shown. We plot also the scaling function (thick dashed line), valence scaling function (thin dashed line) and  resonance fit, given by Eq. 9 of Ref.~\cite{Niculescu:2000tk}, (dotted line). In the right panel the plots of the ratios Eqs.~\ref{ratio_2}) and \ref{ratio_res}, computed for  1.15<W<3.9~GeV$^2$, are shown (solid and dashed lines respectively).  The results are compared with the data from \cite{Niculescu:2000tk}. \label{fig_ep2}}
\end{figure}

\section{Theoretical framework}

In order to compute the resonance structure functions we apply an updated \cite{Graczyk:2007bc} Rein-Sehgal (RS) model \cite{RS}. The RS model \cite{RS} was devoted to describe the $1\pi$ production induced by neutrino-nucleon interaction in a wide kinematical region (up to $W=2.0$~GeV). The description is based on the relativistic harmonic oscillator quark model \cite{FKR}.  Even thought the RS model is relatively simple and old it
is still the reference description for the experimental data analysis in modern experiments like K2K and MiniBooNE.

In the RS model the vector and axial hadronic currents are described with the nucleon vector and axial elastic form factors but, as it was shown in Ref. \cite{Graczyk:2007bc}, applying these form factors leads to the underestimation of  the vector part and overestimation of the axial contribution. In this analysis we use the vector and axial form factors proposed in Ref. \cite{Graczyk:2007bc}. The updated vector contribution agrees with the recent fits of the resonance-like electroproduction data (in the $P_{33}(1232)$ resonance region), while the new axial form factor was obtained as a simultaneous fit to the ANL and BNL $1\pi$ production data\footnote{The axial form factor is given by Eq.~52 from the erratum of Ref.~\cite{Graczyk:2007bc}.}. The updated RS model works efficiently and fits very well to the recent MiniBooNE measurements \cite{Nowak:2009se}.

The mechanism for the duality appearance in lepton scattering off nucleons is partially explained by quark model arguments \cite{quark_model}. The RS model as an example of a quark model with strong experimental applications seems to be an interesting framework for phenomenological studies of the duality in $\nu N$ scattering.

Close \& Isgur claimed that one of the necessary conditions for the realization  of the duality is taking into consideration at least one complete set of resonances of each symmetry-type. Thus, for completeness of our analysis we
consider the following list of resonances\footnote{The interference terms between resonances are also included in the calculus.}:
$P_{33}(1232)$,
$P_{11}(1440)$,
$D_{13}(1520)$,
$S_{11}(1535)$,
$S_{31}(1620)$,
$S_{11}(1650)$,
$P_{33}(1600)$,
$D_{13}(1700)$,
$D_{15}(1675)$,
$F_{15}(1680)$,
$D_{33}(1700)$,
$P_{11}(1710)$,
$P_{13}(1720)$,
$P_{31}(1750)$,
$P_{13}(1900)$,
$S_{31}(1900)$,
$P_{31}(1910)$, 
$F_{35}(1905)$, 
$F_{37}(1950)$, 
$P_{33}(1920)$, 
$F_{17}(1990)$,
$F_{15}(2000)$, 
$F_{35}(2000)$, 
$G_{17}(2190)$, 
$G_{19}(2250)$. However,  one needs to remember that for larger values of the hadronic invariant mass the harmonic oscillator quark model predictions deviate from the experimental measurements. Therefore extending the number of resonances in the discussion can give only a hint on understanding the duality.

In order to quantitatively study the BG duality we introduce the ratio:
\begin{equation}
\label{ratio_def} R_i\left(f_i,Q^2_{res},Q^2_{DIS}\right)=\frac{\displaystyle \displaystyle
\int_{\xi_{min}}^{\xi_{max}} d \xi\, f^{res}(\xi, Q^2_{res})
}{\displaystyle\int_{\xi_{min}}^{\xi_{max}} d \xi\, f^{DIS} (\xi,
Q^2_{DIS})},
\end{equation}
where the limits of integrals are defined by (\ref{limits}),
while  $f_i^{DIS}$ and $f_i^{res}$ are computed with GRV94 PDF's \cite{GRV94}
and  the updated RS model respectively.

\section{$F_2^{ep}$ structure function}

In order to test our theoretical model we compare the theoretical predictions for inclusive $F_2^{ep}$ structure function with the resonance CLAS data~\cite{Osipenko:2003jb}. As can be seen in Fig.~\ref{fig_ep1} the quark model predictions are systematically below the data but it is not a surprise, because the nonresonant background (NoBG) contribution is not taken into account.  On the other hand, the experimental data, and theoretical predictions seem to lie along the valence $F_2^{ep,val}$ curve
(see  Fig.~\ref{fig_ep2}). This property of the resonance data has been already studied in Ref. \cite{Niculescu:2000tj}.

The quantitative verification of our approach is done by computing the ratio (see Fig.~\ref{fig_ep1})
\begin{equation}
\label{ratio_2}
R_2^{ep}(Q^2_{res}) \equiv R\left(F_2^{ep},Q^2_{res},Q^2_{DIS}=10~\mathrm{GeV}^2\right),
\end{equation}
where we consider five different values of $\xi_{min}$, namely  $W_{max}=$1.6, 1.8, 2 and 2.2~GeV. It is seen that,
above $Q^2_{res}=0.5$, the ratios (\ref{ratio_2}) computed for $W_{max} = 2$ GeV and $W_{max} = 2.2$ GeV
slowly vary on $Q^2_{res}$ ($R_2^{ep}(Q^2_{res}>0.5)\sim 0.6$ and $R_2^{ep}(Q^2_{res}>0.5)\sim 0.7$ respectively).  The weak $Q^2_{res}$ dependence of $R_2^{ep}(Q^2_{res})$ is evidence for the same scaling behavior of the resonance predictions and the DIS structure function, however, it is clear that some strength is missing to get a perfect duality.

The scaling is broken below $Q^2_{res}=0.5$~GeV$^2$, where $R^{ep}_2(Q_{res}^{2})$ rapidly tends to zero\footnote{Gauge invariance implies that as $F_2^{ep,res}(Q^2_{res}\to 0)\to 0$; as a consequence, in the ratio of Eq.~\ref{ratio_2},  the numerator decreases faster than the denominator.}.  The argument which explains the mechanism of the duality violation (below $Q^2_{res}=0.5$~GeV$^2$) was given in \cite{Close:2001ha}: in the quark models the duality is strongly related to the dominance of the magnetic contribution, while at low $Q^2$, the electric and magnetic multipoles have comparable strengths.

In order to compare more qualitatively with the experimental measurements  we compute the ratio (Eq.~\ref{ratio_2}) for
$1.1<W^2<3.9$~GeV$^2$ and compare it with the analogous one obtained by Niculescu \textit{et al.} \cite{Niculescu:2000tk}
(here the DIS contribution was predicted with MRS(G) LQ PDF's). Both ratios are plotted in right panel of Fig. \ref{fig_ep2}.
Our result is systematically shifted down with respect to the experimental data.

In Ref. \cite{Niculescu:2000tk} it was shown that if one assumes the BG duality for $F_2^{ep}$ structure function, then  the resonance data can be approximated by the DIS-like $F^{res-par}_2$ parametrization.
The obtained fit (on an average level) describes well resonance data. We use this parametrization in our analysis. In Fig.~\ref{fig_ep2} we plot the ratio
\begin{equation}
\label{ratio_res}
R_2^{res-par}\left(Q^2_{res}\right)=\frac{\displaystyle \displaystyle
\int_{\xi_{min}}^{\xi_{max}} d \xi\, F^{ep,RS}_2(\xi, Q^2_{res})
}{\displaystyle\int_{\xi_{min}}^{\xi_{max}} d \xi\, F^{res-par}_2 (\xi)},
\end{equation}
computed for $1.1<W^2<3.9$~GeV$^2$. We see that the above function weakly depends on $Q^2_{res}$. Additionally, in almost all considered domains $R_2^{res-par}\left(Q^2_{res}\right)$ is around 0.6. It illustrates the fact that the quark model systematically underestimates the experimental data, and it needs to be enriched by the nonresonant background contribution.  On the other hand this regular deviation between quark model predictions and the data allows us to propose the effective structure function as:
\begin{eqnarray}
\label{new_F2}
 F_{2,dual}^{ep}(\xi) &=& F_2^{ep, res}(\xi) + 0.4 \eta(\xi) F_2^{ep, val}, \\
 \eta(\xi) &=& 2\arctan( (W-M -m_\pi)/GeV )/\pi,
\end{eqnarray}
where the $F_2^{ep, val}$ is the valence DIS contribution, while $\eta(\xi)$ function is introduced to make a smooth transition from the threshold to the $P_{33}(1232)$ resonance peak. It is clear the  ratio $R\left(F_{2,dual}^{ep},Q^2_{res},Q^2_{DIS}=10~\mathrm{GeV}^2\right)$  and the analogously computed ratio (Eq.~\ref{ratio_res}) will tend to one.  In Fig.~\ref{fig_ep2} (left panel), we plot  $F_{2,dual}^{ep}$ together with the CLAS data. The good agreement in the first and also second resonance region can be seen.

The above approximation of the structure function seems to be naive in simplicity but very effective in the applications. In particular, it can be useful to apply to  the Monte Carlo generators, dedicated to simulate neutrino-matter interactions.

\section{Neutrino-nucleon interactions}

It is well known that if the $P_{33}(1232)$ resonance region is included in the discussion  then the BG duality can not be simultaneously observed for the proton and neutron targets. It is the result of the isospin symmetry of the resonance structure function and $SU(6)$ symmetry property of the DIS structure function. Fortunately the BG duality (assuming some accuracy) appears for the weak structure functions of the isoscalar target .

For the study of $\nu N$ scattering we apply the updated RS model but the description is supplemented by including the nonresonant background contribution (we follow the effective description proposed in Ref. \cite{RS}).

\begin{figure}
  \includegraphics[width=1.0\textwidth]{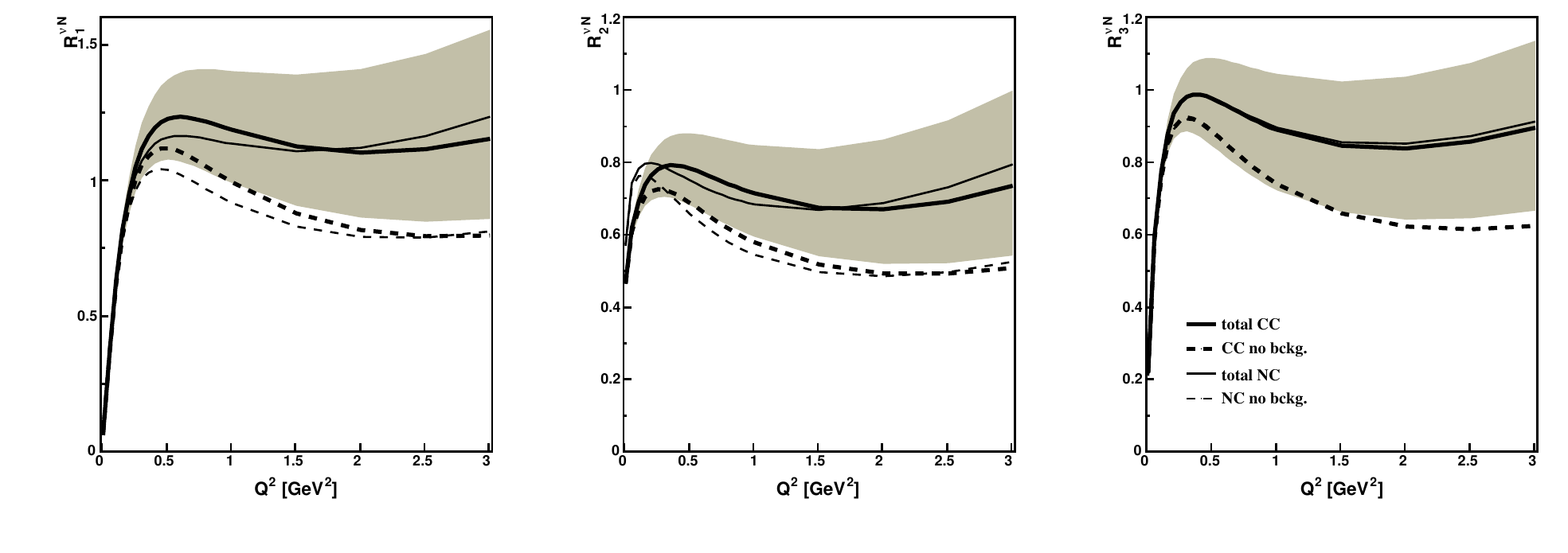}
  \caption{Plots of Eqs.~\ref{ratio_1_nu}, \ref{ratio_2_nu}, and \ref{ratio_3_nu}, computed
  with $W_{max}=1.8$~GeV are shown. The solid lines denote the full model predictions with NoBG.
  The computations without NoBG contribution are denoted by dashed lines. The thick and thin lines
  denote the ratios for CC and NC  structure functions. The shadow areas denote the $1\sigma$ uncertainty.
  \label{fig_neutrino}}
\end{figure}

We consider the weak charged current (CC) and neutral current (NC) nucleon structure functions. The following ratios are computed:
\begin{eqnarray}
\label{ratio_1_nu}
R_1^{\nu N}(Q^2_{res})& \!\!\equiv\!\!
& \!\! R\left(xF_1^{\nu N},Q^2_{res},Q^2_{DIS}=10~\mathrm{GeV}^{2}\right),  \\
\label{ratio_2_nu}
R_2^{\nu N}(Q^2_{res})& \!\!\equiv\!\!
& \!\! R\left(F_2^{\nu N},Q^2_{res},Q^2_{DIS}=10~\mathrm{GeV}^{2}\right),  \\
R_3^{\nu N}(Q^2_{res})& \!\!\equiv\!\! & \!\! R\left(x
F_{3}^{\nu N},Q^2_{res},Q^2_{DIS}=10~\mathrm{GeV}^{2}\right)\;.
\label{ratio_3_nu}
\end{eqnarray}
We found out that for $W_{max}=1.8$~GeV (see Fig. \ref{fig_neutrino}) the above ratios, similarly as for $ep$ scattering, weakly depend on $Q^2_{res}$. Once again  it can be treated as evidence for appearance of the duality.

It is interesting that the ratios of CC and NC structure functions are very similar. It is the result of the $SU(3)\times SU(2)$ and $SU(6)$ symmetries  of the structure functions. As a consequence if the duality is observed for the one channel, say CC, it must be also visible (on the same level) for the NC structure functions. This property  can have applications in modeling of the cross sections for the neutral current neutrino-nucleus scattering.

As was mentioned in first section, the predictions of the cross sections for neutrino-nucleon scattering are affected by the lack of knowledge of the axial structure of the nucleon. Therefore
we present the ratios (\ref{ratio_1_nu}-\ref{ratio_3_nu}) with $1\sigma$ error coming from the axial form factor uncertainty  (we apply the results of  Ref. \cite{Graczyk:2009qm}, where the uncertainty of the axial contribution was discussed). It is seen (in Fig.~\ref{fig_neutrino}) that for the $xF_3^{\nu N}$ and $xF_1^{\nu N}$ structure functions the BG duality is observed within $1\sigma$ error, whereas for $F_2^{\nu N}$ structure function  the duality appears within $2\sigma$ error.

However, we remark that in the discussion of structure function uncertainties, one needs to remember that in our approach the description of  NoBG contribution is quite approximate. It  gives rise to a systematic error. In order to show the role of NoBG contribution we plot the ratios (Eqs.~\ref{ratio_1_nu}--\ref{ratio_3_nu}) computed  without NoBG.

\section{Outlook}

As was mentioned in the first section, the duality is observed in electron-deuteron scattering.
But the scaling behavior of the resonance data for electron scattering off heavier nuclear targets,
like carbon and iron \cite{Arrington:2003nt}, is also observed. Here the duality between the resonance
and the DIS nuclear structure function is visible even before averaging the data.

In fact, the Fermi motion effect naturally ''averages''  the nucleon structure function over a wide  kinematical region, but,  there are also other nuclear medium effects like renormalization of the resonance properties, redistribution of the resonance prominent regions etc.. It makes the theoretical investigation of the duality in lepton scattering off nuclei very difficult and delicate.

On the other hand, observing the duality for nuclear targets means that the nuclear corrections in the resonance and the DIS regions are comparable. But it has been not expected \cite{Melnitchouk:2005zr}.

Some effort to investigate the duality in lepton scattering off nuclei was done by Lalakulich \textit{et al.} \cite{Lalakulich:2008tu}. The electromagnetic and weak nuclear structure functions
were obtained with an independent shell particle model, and then compared with the experimental data. For the $\nu$-nuclei scattering the duality was not obtained. However, probably further studies of the nuclear effects and the role of NoBG contribution can shed light on this topic \cite{Leitner:2009de}.

It is obvious that assuming that the Bloom-Gilman duality is also a fundamental property of weak nuclear structure
functions can give interesting applications in neutrino scattering physics. In particular in can be useful for:
\begin{itemize}
\item[(i)]  modeling the neutrino cross sections in the resonance region;
\item[(ii)]  extraction of the axial contribution from the scattering data;
\item[(iii)] studying the nonresonant background contribution.
\end{itemize}
Hopefully the forthcoming neutrino experiments, like Miner$\nu$a will be able to critically investigate the duality in neutrino scattering off nuclei.


\begin{theacknowledgments}
The author was supported by the grant:  35/N-T2K/2007/0 (the project number DWM/57/T2K/2007).
\end{theacknowledgments}


\end{document}